\documentclass[12pt]{article}

\newcommand{\eqn}[1]{(\ref{#1})}
\newcommand\pubnumber{}
\newcommand\pubdate{December 22 2001}
\newcommand\hepnumber{hep-ph/0112291}

\def\csumb{$^a$Department of Physics, East China Normal University,\\
        \em Shanghai 200062, P R China}
\def\support{\footnote{Work supported in part by National Nature Science
Foundation of China.}}

\def\Title#1{\begin{center} {\Large\bf #1 } \end{center}}
\def\Author#1{\begin{center}{ \sc #1} \end{center}}
\def\Address#1{\begin{center}{ \it #1} \end{center}}

\newcommand\pubblock{\rightline{\begin{tabular}{l} \pubnumber\\
         \pubdate\\ \hepnumber \end{tabular}}}
\newenvironment{Abstract}{\begin{quotation}  }{\end{quotation}}

\makeatletter
\def\section{\@startsection{section}{0}{\z@}{5.5ex plus .5ex minus
 1.5ex}{2.3ex plus .2ex}{\large\bf}}
\def\subsection{\@startsection{subsection}{1}{\z@}{3.5ex plus .5ex minus
 1.5ex}{1.3ex plus .2ex}{\normalsize\bf}}
\def\subsubsection{\@startsection{subsubsection}{2}{\z@}{-3.5ex plus
-1ex minus  -.2ex}{2.3ex plus .2ex}{\normalsize\sl}}

\renewcommand{\@makecaption}[2]{%
   \vskip 10pt
   \setbox\@tempboxa\hbox{\small #1: #2}
   \ifdim \wd\@tempboxa >\hsize     
       \small #1: #2\par     
     \else                        
       \hbox to\hsize{\hfil\box\@tempboxa\hfil}
   \fi}

 \def\citenum#1{{\def\@cite##1##2{##1}\cite{#1}}}
\newcount\@tempcntc
\def\@citex[#1]#2{\if@filesw\immediate\write\@auxout{\string\citation{#2}}\fi
  \@tempcnta\z@\@tempcntb\m@ne\def\@citea{}\@cite{\@for\@citeb:=#2\do
    {\@ifundefined
       {b@\@citeb}{\@citeo\@tempcntb\m@ne\@citea\def\@citea{,}{\bf ?}\@warning
       {Citation `\@citeb' on page \thepage \space undefined}}%
    {\setbox\z@\hbox{\global\@tempcntc0\csname b@\@citeb\endcsname\relax}%
     \ifnum\@tempcntc=\z@ \@citeo\@tempcntb\m@ne
       \@citea\def\@citea{,}\hbox{\csname b@\@citeb\endcsname}%
     \else
      \advance\@tempcntb\@ne
      \ifnum\@tempcntb=\@tempcntc
      \else\advance\@tempcntb\m@ne\@citeo
      \@tempcnta\@tempcntc\@tempcntb\@tempcntc\fi\fi}}\@citeo}{#1}}
\def\@citeo{\ifnum\@tempcnta>\@tempcntb\else\@citea\def\@citea{,}%
  \ifnum\@tempcnta=\@tempcntb\the\@tempcnta\else
  {\advance\@tempcnta\@ne\ifnum\@tempcnta=\@tempcntb \else\def\@citea{--}\fi
    \advance\@tempcnta\m@ne\the\@tempcnta\@citea\the\@tempcntb}\fi\fi}
\makeatother

\begin{document}
\begin{titlepage}
\pubblock \vfill
\def\thefootnote{\fnsymbol{footnote}}
\Title{Dynamical symmetry breaking of massless $\lambda\phi^4$ and
renormalization scheme dependence\support} \vfill \Author{Ji-Feng
Yang$^{a}$} \Address{\csumb}

\begin{Abstract}
Through two loop effective potential we demonstrate that the
$\overline{MS}$ scheme of dimensional regularization and Jackiw's
prescription in cut-off regularization allow for the dynamical
breaking solution in massless $\lambda\phi^4$ theory, while the
Coleman-Weinberg prescription does not. The beta function of the
broken phase is negative, like in the one loop effective
potential, but the UV fixed point is not zero, i.e., not an
asymptotic freedom solution, unlike the one loop case. Some
related issues were briefly discussed. \\

\end{Abstract}
\vfill
\end{titlepage}
\def\thefootnote{\arabic{footnote}}
\setcounter{footnote}{0}

\section{Introduction}
In spite of the magnificent success of the standard model, the
genuine symmetry breaking mechanism remains elusive as the Higgs
sector is unsatisfactory due to hierarchy\cite{Hie},
triviality\cite{Trivial}and other problems. In the pursuit of more
satisfactory models for symmetry breaking, supersymmetric field
theories\cite{SUSY} (including string theories and their
descendants\cite{String}) and technicolor model and its
variants\cite{TC} have been intensively studied. Although these
models already differ among themselves, they share a common
feature: the elementary Higgs scalar fields are excluded. And it
is widely believed that the solution to the hierarchy and
triviality problem must be in non-perturbative regime\cite{SUSY}.

On the other hand there are also some efforts to revive the
$\lambda\phi^4$ interaction through the asymptotic free
renormalization\cite{Stevenson,Branchina} proposal, i.e.,
$\beta(\lambda)<0$ with $\lambda=0$ as the UV fixed point, in
contrast to the perturbative renormalization where
$\beta(\lambda)>0$ (leading to triviality), i.e., the
renormalization prescription is physically relevant.

In this report we wish to investigate the existence of the
dynamical symmetry breaking in the two loop effective potential.
For simplicity, we will consider the simplest scalar model, the
massless $\lambda\phi^4$ model with $Z_2$ symmetry, the first
example with which the dynamical symmetry breaking was
demonstrated\cite{CW}. There is also a technical concern in
choosing massless scalar theory: there is no non-convexity
associated with the tachyon mass term in massless models unlike
the Higgs model\cite{Convex} and the configuration of the
expectation value of scalar field automatically becomes
homogeneous. In the meantime we also wish to examine the relevance
of renormalization prescription.

The paper is organized as follows. The bare and renormalized two
loop effective potentials will be given in several prescriptions
in section two. Then in section three the symmetry breaking
solution is shown to exist in several renormalization
prescriptions but not in the Coleman-Weinberg\cite{CW}
prescription. Some discussions and the summary will comprise the
last section.
\section{The intermediate renormalization of the two loop
effective potential of massless $\lambda\phi^4$} As is explained
in the introduction, we consider the massless $\lambda\phi^4$
model with $Z_2$ symmetry: invariance under the transformation of
$\phi\rightarrow -\phi$. The algorithm for two loop effective
potential is well known according to Jackiw\cite{Jackiw}
\begin{eqnarray}
\textsc{L}&=&\frac{1}{2}(\partial\phi)^2-\lambda\phi^4,\\
V_{(2l)}&\equiv& \lambda\phi^4+\frac{1}{2}I_{0}(\Omega)+3\lambda
I^2_1(\Omega)-48\lambda^2 \phi^2I_2(\Omega),\\
\Omega&\equiv& \sqrt{12\lambda\phi^2};\\
I_{0}(\Omega)&=&\int\frac{d^4 k}{(2\pi)^4}\ln(1+\frac{\Omega^2}{k^2});\\
I_1(\Omega)&=&\int\frac{d^4 k}{(2\pi)^4}\frac{1}{k^2+\Omega^2};\\
I_2(\Omega)&=&\int\frac{d^4 k d^4
l}{(2\pi)^8}\frac{1}{(k^2+\Omega^2)(l^2+\Omega^2)((k+l)^2+\Omega^2)};
\end{eqnarray}
Here we have Wick rotated all the loop integrals into Euclidean
ones. As these integrals have already been calculated in
literature both in dimensional regularization\cite{FJ} and in
cut-off scheme\cite{Jackiw}, we only list the results here. The
bare forms in the two regularization schemes read:
\begin{eqnarray}
 V^{(D) }_{(2l)}( \Omega ) &=&\Omega ^{4}\{
\frac{1}{144\lambda }+\frac{-\frac{1}{\epsilon }
+\overline{L}-\frac{3}{2}}{ ( 8\pi ) ^{2}}+\frac{3\lambda }{( 4\pi
) ^{4}}[ (-\frac{1}{\epsilon }+\overline{L}-1) ^{2}+(
\overline{L}-1)^{2}\nonumber \\
&&+2( -\frac{1}{\epsilon }+\overline{L}-\frac{3}{2}) ^{2}+2(
\overline{L}-\frac{3}{2}) ^{2}+7+6S-\frac{5}{3}\zeta(2)] \};\\
V^{(\Lambda )}_{(2l)}(\Omega )&=&\Omega ^{4}\{ \frac{1}{144\lambda
}+\frac{ L_{\Lambda }^{\Omega }-\frac{1}{2}}{( 8\pi )
^{2}}+\frac{3\lambda }{( 4\pi ) ^{4}}[ (L_{\Lambda }^{\Omega
})^{2}-2+2( L_{\Lambda
}^{\Omega }-1) ^{2}] \} \nonumber \\
&&+\Omega ^{2}\{ \frac{2\Lambda ^{2}}{(8\pi )^{2}}+\frac{3\lambda
\Lambda ^{2}L_{\Lambda }^{\Omega }}{(4\pi )^{4}}-\frac{8\lambda
\Lambda ^{2}}{(4\pi )^{4}}\}.
\end{eqnarray}where $S=\sum^{\infty}_{n=0}\frac{1}{(2+3n)^2},
\overline{L}=L+\gamma-\ln 4\pi$ , $L=\ln \frac{\Omega^2}{\mu^2}$
and $L^{\Omega}_{\Lambda}=\ln \frac{\Omega^2}{\Lambda^2}$. Here we
have omitted all the field independent terms.

In literature the renormalization had been done in the
$\overline{MS}$ scheme for $V^{(D) }_{(2l)}( \Omega )$
(Cf.\cite{FJ}), while for $V^{(\Lambda )}_{(2l)}(\Omega )$ the
renormalization had been done in Jackiw's
prescription\cite{Jackiw} and Coleman-Weinberg's
prescription\cite{CW}. The results read
\begin{eqnarray}
V^{(\overline{MS}) }_{(2l)}( \Omega )
&=&\Omega^{4}\{\frac{1}{144\lambda}+\frac{\overline{L}
-\frac{3}{2}}{(8\pi)^2}+\frac{3\lambda}{(4\pi)^4}[3\overline{L}^2
-10\overline{L}+11+12S-\frac{8}{9}\pi^2]\};\\
V^{(Jackiw)}_{(2l)}(\Omega )
&=&\Omega^{4}\{\frac{1}{144\lambda}+\frac{\check{L}}{(8\pi)^2}+
\frac{3\lambda}{(4\pi)^4}[3\check{L}^2-\check{L}]\};\\
V^{(CW)}_{(2l)}(\Omega )
&=&\Omega^{4}\{\frac{1}{144\lambda}+\frac{\breve{L}}{(8\pi)^2}+
\frac{3\lambda}{(4\pi)^4}[3\breve{L}^2-\breve{L}+\frac{205}{12}]\},
\end{eqnarray}with $\check{L}=\ln
\frac{\Omega^2}{12\lambda\mu^2_{Jackiw}}$ and $\breve{L}=\ln
\frac{\Omega^2}{12\lambda\mu^2_{CW}}-\frac{25}{6}$. In all the
above formulae the scheme (or prescription) dependence of field
strength and coupling constant are understood. Note that the
Jackiw and Coleman-Weinberg prescriptions were applied to the same
bare effective potential, i.e., that calculated in the cut-off
scheme.

Upon appropriate rescaling of the subtraction scales, all versions
of the effective potential take the following form (we will drop
all the dressing symbols)
\begin{eqnarray}
\label{SCH} V_{(2l)}( \Omega )
=\Omega^{4}\{\frac{1}{144\lambda}+\frac{L-1/2}{(8\pi)^2}
+\frac{3\lambda}{(4\pi)^4}[ L^2+2(L-1)^2+\alpha]\}
\end{eqnarray}with $L \equiv \ln \frac{\Omega^2}{\mu^2}$. Now we can
see the explicit scheme dependence represented by the constant
$\alpha$ that is summarized in the following tabular:

\begin{tabular}{|c|c|}
   \hline
    Value of $\alpha$  &  Reg and/or Ren Scheme\\
   \hline
  $-2.6878$   & $\overline{MS}$ \\
    \hline
  $-\frac{5}{4} $ & Jackiw \\
  \hline
  $16\frac{1}{3} $ & Coleman-Weinberg
   \\ \hline
\end{tabular}

Such scheme dependence of the effective potential differs from the
case of truncating the perturbative scattering
matrix\cite{scheme}\footnote{Rigorously speaking, this property
has been firmly established only in massless gauge theories or in
cases where threshold effects are negligible. But the relevance of
renormalization prescriptions has been recently emphasized in
theories with unstable elementary particles (like $W^{\pm},Z^0$
bosons in electro-weak theory)\cite{Pole}.}: The difference in
$\alpha$ could not be removed by redefinition of coupling constant
(and perhaps of field strength) without changing the functional
form of the effective potential.
\section{Effective potential and symmetry breaking}
Now let us look at the dynamical symmetry breaking. Working with
the general parametrization form of Eq.~\eqn{SCH}, the task is to
solve the following equation:
\begin{eqnarray}
\label{SCH2}
\frac{d V_{(2l)}( \sqrt{12\lambda\phi^2} )}{d\phi}=
24\lambda\phi\Omega^2[\frac{2V_{(2l)}(\Omega^2)}{\Omega^4}
+\frac{1}{(8\pi)^2}+\frac{3\lambda}{(4\pi)^4}(6L-4)]=0.
\end{eqnarray}

An obvious solution is $\phi=0$ which is the symmetric solution in
perturbative (weak coupling) regime while the symmetry breaking
solution is determined by the following algebraic equation
\begin{eqnarray}
\label{SCH3}
3L^2+(\frac{4\pi^2}{3\lambda}-1)L+\alpha+\frac{16\pi^4}{27\lambda^2}=0.
\end{eqnarray} Here it is obvious that existence of real number
solution depends on both $\alpha$ and $\lambda$.
\subsection{Determinants of symmetry breaking}
Since we must start from a stable micro potential the coupling
$\lambda$ must be a positive real number. Now let us closely
examine Eq.~\eqn{SCH3}. For Eq.~\eqn{SCH3} to possess a finite
real number solution, we must impose the following inequality in
terms of $\alpha$ and $\lambda$
\begin{eqnarray}
\label{Delta}
\Delta\equiv
(\frac{4\pi^2}{3\lambda}-1)^2-12(\alpha+\frac{16\pi^4}{27\lambda^2})=
\frac{1}{3}[4-36\alpha-(1+\frac{4\pi^2}{\lambda})^2] \geq 0.
\end{eqnarray}
This inequality is only valid for certain ranges of $\alpha$ and
$\lambda$,
\begin{eqnarray}
\label{alpha}
\alpha &<& \frac{1}{12},\\
\lambda&\geq&
\lambda_{cr}\equiv\frac{4\pi^2}{\sqrt{4-36\alpha}-1}.
\end{eqnarray}

Then the solutions to Eq.~\eqn{SCH3} can be found with the above
two requirements,
\begin{eqnarray}
L_{\pm}(\lambda)=\frac{1}{6}[1-\frac{4\pi^2}{3\lambda}\pm
\sqrt{\Delta}],
\end{eqnarray}from which and the definitions $L \equiv \ln
\frac{\Omega^2}{\mu^2}, \Omega \equiv \sqrt{12\lambda\phi^2}$ we
can find the nonzero solutions of $\phi$ which read
\begin{eqnarray}
\phi^2_{\pm}(\lambda;[\mu,\alpha])=\frac{\mu^2}{12\lambda}
\exp\{\frac{1}{6}[1-\frac{4\pi^2}{3\lambda}\pm\sqrt{\Delta}]\}.
\end{eqnarray}
But the solutions corresponding to $L_{-}(\lambda)$ are local maxima
(tachyonic), only the $ L_{+}(\lambda)$ solutions are local minima,
this can be seen from the second order derivative of the effective
potential at $\Omega^2_{\pm}$ (which is exactly the effective mass),
\begin{eqnarray}
\label{mass}
m_{eff;\pm}(\lambda)\equiv \frac{\partial^2 V_{(2l)}}{(\partial\phi)^2}
\|_{\phi^2=\phi_{\pm}}=\pm \frac{18\lambda^2\Omega^2_{\pm}}{(2\pi)^4}
\sqrt{\Delta},
\end{eqnarray}as each factor is positive definite.

The inequality~\eqn{alpha} shows the relevance of renormalization
prescriptions at two loop level: the Coleman-Weinberg prescription
could not lead to dynamical symmetry breaking at all as the
critical inequality~\eqn{alpha} is violated:
$\alpha_{CW}=16\frac{1}{3}=\frac{196}{12} \gg \frac{1}{12}$,
similar to the one loop case\cite{Stevenson,Branchina}.
\subsection{Stability of symmetry breaking and criterion for
coupling constant} Using Eq.~\eqn{SCH2} we have
\begin{eqnarray}
\label{vac} E_{\pm}(\lambda,\mu)\equiv
V_{(2l)}(\sqrt{12\lambda\phi^2})|_{\phi^2=\phi_{\pm}^2}=
-\frac{(12\lambda\phi^2_{\pm})^2}{2(8\pi)^2}
[1+\frac{3\lambda}{2\pi^2}(3L_{\pm}-2)].
\end{eqnarray}For the symmetry breaking states to be stable,
the vacuum energy density must be less than zero. Then from
Eq.~\eqn{vac} we find that,
\begin{eqnarray}
\label{stablecondition} L_{\pm} \geq
\frac{2}{3}-\frac{2\pi^2}{9\lambda},
\end{eqnarray}which rejects the $L_{-}$ solution and leads to
the following requirement in $L_{+}$,
\begin{eqnarray}
\lambda\geq \hat{\lambda}_{cr}\equiv
\frac{4\pi^2}{\sqrt{4-36\alpha-27}-1}(>\lambda_{cr}=
\frac{4\pi^2}{\sqrt{4-36\alpha}-1}).
\end{eqnarray} The values of the
critical couplings are exhibited in the following tabular.

\begin{tabular}{|c|c|c|}
\hline
  Scheme & $\lambda_{cr}$ & $\hat{\lambda}_{cr}$ \\
  \hline
  $\overline{MS}$ & 4.368& $5.2024$ \\
    \hline
  Jackiw & 6.5797 & 10.698 \\ \hline
\end{tabular}

We also note that the stable condition~\eqn{stablecondition}
amounts to the following more stringent requirement on
renormalization prescriptions:
\begin{eqnarray}
\alpha\leq-\frac{2}{3}.
\end{eqnarray}
\subsection{RG invariance of vacuum energy and beta function}
Since the vacuum energy is a physical entity, it must be
renormalization group invariant, i.e., insensitive to the choice
of subtraction point within a scheme\cite{Branchina}. Therefore we have
\begin{eqnarray}
\label{RGI} \mu\frac{d E_{+}(\lambda,\mu)}{d\mu}=0.
\end{eqnarray}We must stress that this important condition in fact
defines a fundamental physical scale as input, corresponding to
the necessary step after intermediate renormalization is done,
i.e., to confront the renormalized amplitudes with experiments or
to fix the amplitudes via physical conditions.

From this equation we can determine the beta function of $\lambda$
as was did in ref.\cite{Branchina}. First let us rewrite the
vacuum energy density as
\begin{eqnarray}
\label{vac2}
E_{+}=-\frac{\mu^4}{2(8\pi)^2}\varepsilon_{+}
(\lambda)e^{L_{+}(\lambda)},
\end{eqnarray}with
\begin{eqnarray}
\label{epsilon} \varepsilon_{+}(\lambda)\equiv1+
\frac{3\lambda}{2\pi^2}(3L_{+}(\lambda)-2)=\frac{3\lambda}{4\pi^2}
(\sqrt{\Delta}-3). \end{eqnarray} Then we find from Eqs.~\eqn{RGI}
and ~\eqn{vac2} that,
\begin{eqnarray}
\beta(\lambda)=-\frac{12\lambda \varepsilon_{+}(\lambda)}
{\{\varepsilon_{+}(\lambda)(3+\frac{4\pi^2}{3\lambda})+
1+4\pi^2/\lambda\}}<0.
\end{eqnarray}

This is true for all the two schemes allowing for symmetry
breaking solution. When the coupling becomes infinitely strong,
i.e., $\lambda\rightarrow\infty$, the beta function approaches to
a straight line:
\begin{eqnarray}
\beta(\lambda)|_{\lambda\rightarrow\infty}\sim -4\lambda,
\end{eqnarray}while in the neighborhood of $\hat{\lambda}_{cr}$,
\begin{eqnarray}
\beta(\lambda)|_{\lambda\rightarrow \hat{\lambda}^{+}_{cr}}\sim
-4(\lambda-\hat{\lambda}_{cr}).
\end{eqnarray}

Our faith in the two loop effective potential is enhanced by the
fact that all schemes (except the Coleman-Weinberg prescription)
predict the same kind of running behavior of the coupling. As a
rude approximation we use the following beta function,
\begin{eqnarray}
\label{rude}
\beta_{appr}(\lambda)=-4(\lambda-\hat{\lambda}_{cr}),\forall\lambda:
\lambda\geq\hat{\lambda}_{cr}
\end{eqnarray}with the obvious solution
\begin{eqnarray}
\label{rude2}
\lambda-\hat{\lambda}_{cr}=\frac{\mu^4_0}{\mu^4},\forall\mu:
\mu\in(0,\infty)
\end{eqnarray}which could also be obtained as a rude approximation
of Eq.~\eqn{vac}. In fact the RG invariant scale $\mu^2_{0}$ in
the IR end differs from that in the UV end, but this fact does not
affect the main properties described by the
approximation~\eqn{rude} and~\eqn{rude2}.

Now it is clear that we obtained a \emph{nontrivial} theory with a
UV fixed point equal to the value of critical
coupling,$\hat{\lambda}_{cr}$, a strong coupling as is clear from
the tabular in last subsection. From Eq.~\eqn{rude2} we can
identify a pole in the IR region, which is quite different from
the IR Landau pole in QCD. Since the coupling is strong in the
entire broken phase, we found a solution entirely lives in strong
coupling regime with the IR end being infinitely strongly coupled!
Note that this property is true in several schemes (except the
Coleman-Weinberg prescription) such as $\overline{MS}$. In the
following discussions, we refer to this solution as SCRDSB for
Strong Coupling Regime Dynamical Symmetry Breaking.

The asymptotic behaviors of the effective mass defined in
Eq.~\eqn{mass} can be obtained after some calculations (taking the
vacuum energy density as fundamental physical quantity)
\begin{eqnarray}
m^2_{eff}(\lambda)\|_{\lambda\sim\infty}\sim
\lambda^{\frac{3}{2}};\ \ \ \ \ \
m^2_{eff}(\lambda)\|_{\lambda\sim\hat{\lambda}_{cr}^{+}}
  \sim(\lambda-\hat{\lambda}_{cr})^{-\frac{1}{2}}.
\end{eqnarray} Or in terms of running scale
\begin{eqnarray}
m^2_{eff}(\mu^2)\|_{\mu\sim0}\sim \frac{1}{\mu^6};\ \ \ \ \ \
m^2_{eff}(\mu^2)\|_{\mu\sim\infty}\sim\mu^2.
\end{eqnarray}

The effective coupling, defined as $\lambda_{eff} (\lambda)\equiv
\frac{\partial^4 V_{(2l)}}{(\partial\phi)^4}
\|_{\phi^2=\phi^2_{+}}$ possesses the following asymptotic
behaviors
\begin{eqnarray}
\lambda_{eff}(\lambda)\|_{\lambda\sim\infty}\sim10^1\lambda^3 ;
\ \ \ \ \ \lambda_{eff}(\lambda)\|_{\lambda\sim\hat{\lambda}_{cr}}
\sim 10^2 \hat{\lambda}_{cr},
\end{eqnarray} or equivalently
\begin{eqnarray}
\lambda_{eff}(\mu)\|_{\mu\sim0}\sim\frac{1}{\mu^{12}};
\ \ \ \ \ \lambda_{eff}(\mu)\|_{\mu\sim\infty}\sim10^3.
\end{eqnarray}Note that the effective coupling becomes more singular
than the effective mass does in the IR limit.

\section{Discussion and summary}
Thus far we just made use of the well known two loop calculations
to search for the symmetry breaking solutions. Our results here
are new in two aspects: (1) First, the scheme dependence of the
non-perturbative framework differs from that of the standard
perturbative framework\cite{scheme}; (2). Second, the broken phase
is a new kind of nontrivial dynamics, i.e., a totally
(non-perturbative) strong coupling dynamics regime with negative
beta function (SCRDSB).

For the first aspect, we note that the standard perturbative
scheme dependence relies on the following well known fact: in the
complete sum of the one particle irreducible (1PI) loop diagrams
at a given order, the real divergence after all sub-divergences
were removed is only a single log divergence, i.e.,
$\sim\ln\Lambda^2$ (or single power divergence,
$\sim\frac{1}{\epsilon}$) in gauge theories like QCD and QED. This
key point then implies that for an experimentally interested
quantity (like a reaction ratio $R_{\cdots}(Q)$), all the energy
scale dependence (via $\ln^n\frac{Q^2}{\mu^2}, n\geq1$) in the
coefficients ($r_i (Q), i\geq1$) of the perturbative expansion of
the quantity in terms of the coupling constant
($R_{\cdots}(Q)=R_{0}(1+\sum^N_{i=1}r_i \alpha^i(\mu))$) can be
absorbed into the energy scale dependence of the coupling constant
through RG improvement\cite{scheme}. This property is ensured by
the gauge invariance of these models.

But the situation is changed in massive $\lambda\phi^4$ due to the
appearance of the double log term in the sum of two loop 1PI
diagrams. In the previously mentioned example\cite{NILOU}, the
relevance of renormalization prescription arises because the
method in use only assumes a subset of diagrams and therefore
fails the redefinition of running parameters (this is also true of
the one loop effective potential case\cite{Stevenson,Branchina}),
here we found that, the problem is technically due to the
interaction structures. Hence, this problem would persist in
higher orders massive diagrams even in \emph{perturbative}
approaches. However, the massive loop diagrams considered here
\emph{are} non-perturbative by construction\cite{Jackiw}, but we
must note that the sum of the 1PI diagrams in the potential is
complete at given order and more importantly the method of
effective potential is systematic. Of course in pure perturbative
framework of massless $\lambda\phi^4$ (without dynamical masses),
there is no double log terms in the sums of the 1PI diagrams and
the perturbative scheme dependence is as usual\cite{PS}.

Now we turn to the possible physical implications of SCRDSB. The
one loop effective potential can be renormalized in
non-perturbative way so that asymptotic freedom ensued
consequently\cite{Stevenson,Branchina}, while here we found a
nontrivial solution entirely living in the strong coupling regime
after symmetry breaking. Although this phenomenon is a two loop
level result, at least there is one thing that is conspicuous: the
two loop solution supports the existence of nontrivial dynamical
symmetry breaking solution that is first discovered one loop case.
We think nontrivial solution might persist after including still
higher order contributions, with the running behaviors might be
more complicated, perhaps with more stringent constraints on the
scheme choices.

Since the entire broken phase is in the strong coupling regime, it
is very difficult to find the elementary scalar fields in
asymptotic final states, as far as the new nontrivial solution is
concerned. This might somehow hint us an alternative possibility
for the Higgs particles' hunting: it might be impossible to detect
the Higgs bosons in the final states in high energy experiments.
Maybe some sort of bound states of the scalar particles could be
found, but such bound states might in turn add difficulties for us
to discern their true origin: from fermion fields condensate (like
in the technicolor model) or from some mysterious scalar fields,
etc. Since the SCRDSB is entirely non-perturbative, it is more
important to see the influence of still higher order contributions
to dynamical symmetry breaking. At this moment it is not clear
what our results here imply to the hierarchy or unnaturalness
problem, but the predictions of the upper and/or lower bounds on
Higgs masses based on perturbative triviality should be reexamined
with the non-triviality solution in mind. It is also interesting
to see how the non-perturbative result here could be applied to
SUSY like contexts. We will make no comment about it before
further study is done.

In summary, we performed a new investigation on the two loop
effective potential of massless $\lambda\phi^4$ and found that the
dynamical symmetry breaking solution entirely lives in the strong
coupling regime. As in the one loop case, there is still
nontrivial relevance of renormalization schemes or prescriptions
at the two loop level. Some related issues were briefly touched.
\section*{Acknowledgement}
The author is grateful to Professor W. Zhu for helpful discussion
and encouragements.

\end{document}